\documentclass[a4paper,11pt]{article}
\pdfoutput=1 

\usepackage{jcappub} 

\usepackage[T1]{fontenc} 
\usepackage{booktabs}
\usepackage{subcaption}
\usepackage{enumitem} 
\usepackage{siunitx}
\usepackage{orcidlink}
\usepackage{makecell}
\graphicspath{{./plots/}}

\newcommand{\revise}{}

\title{Forecasts of constraining isotropic cosmic birefringence on AliCPT-1}
\author[a,b,1]{Jiazheng Dou\,\orcidlink{0000-0001-8024-3931},\note{Corresponding author.}}
\author[a,b,c]{Wen Zhao\,\orcidlink{0000-0002-1330-2329}}
\affiliation[a]{Department of Astronomy, University of Science and Technology of China, Chinese Academy of Sciences, Hefei, Anhui 230026, People’s Republic of China}
\affiliation[b]{School of Astronomy and Space Sciences, University of Science and Technology of China, Hefei 230026, People’s Republic of China}
\affiliation[c]{College of Physics, Guizhou University, Guiyang 550025, People’s Republic of China}

\emailAdd{doujzh@mail.ustc.edu.cn}
\emailAdd{wzhao7@ustc.edu.cn}

\abstract{Cosmic birefringence (CB) is a promising probe of parity-violating physics beyond the Standard Model, characterized by the rotation of the linear polarization plane of cosmic microwave background (CMB) photons. This effect, quantified by the birefringence angle $\beta$, generates non-zero $EB$ and $TB$ correlations that are otherwise absent in standard cosmology. However, instrumental miscalibration angles $\alpha$ can mimic this signal, necessitating a joint estimation approach. In this work, we forecast the sensitivity of the AliCPT experiment, combined with \textit{Planck} HFI data, on constraining the isotropic CB angle using a semi-analytical maximum-likelihood method. We simulate observations under various foreground complexities, rotation angles, and scanning strategies, and demonstrate that AliCPT can achieve an uncertainty of $\sigma(\beta)=0.09^\circ$ with one-year data, which will improve to $0.026^\circ$ after four years' observations. We also find that neglecting or mismodeling the foreground $EB$ correlation will introduce significant biases, which can be alleviated under a clean but small sky patch.
}

\begin{document}
\maketitle
\flushbottom

\section{Introduction}

Cosmic birefringence (CB) has emerged as a unique window into parity-violating physics beyond the Standard Model of cosmology, whose natural result is the rotation of the linear polarization plane of photons as they propagate through space \cite{carrollLimitsLorentzParityviolating1990,komatsuNewPhysicsPolarised2022}. This effect arises from a Chern-Simons interaction coupling a pseudo-scalar field $\phi$ to the electromagnetic tensor $F_{\mu\nu}$ and its dual $\tilde{F}_{\mu\nu}$, via a Chern-Simons term $\mathcal{L}_{\phi\gamma}=-\frac{1}{4}g_{\phi\gamma}\phi F_{\mu\nu}\tilde{F}_{\mu\nu}$ as an addition to the electromagnetic Lagrangian density, where $g_{\phi\gamma}$ is the coupling constant \cite{fengSearchingViolationCosmic2006}. The parity-odd field $\phi$, which changes sign under spatial inversion transformation, represents an undiscovered field beyond the Standard Model framework \cite{marshAxionCosmology2016,Ferreira:2020fam} and a potential candidate for both dark matter \cite{Liu:2006uh,Liu:2016dcg,Nakagawa:2021nme,Obata:2021nql} and dark energy \cite{Choi:2021aze,Fujita:2020ecn,Eskilt:2023nxm,Lin:2025gne}. The detection of a non-zero cosmic birefringence angle would thus constitute direct evidence for parity violation and new physics \cite{pgw3,pgw4,pgw5}.

The cosmic microwave background (CMB) provides the most sensitive observational probe for this phenomenon to date \cite{Lue:1998mq,pgw2}. The rotation angle of the linear polarization plane, commonly referred to as the cosmic birefringence angle $\beta$, accumulates with the propagation distance of photons as $\phi$ varies with time, which is given by $\beta=\frac{1}{2}g_{\phi\gamma}\Delta\phi$, where $\Delta\phi$ is the variation of $\phi$ during the period from emission to observation of photons \cite{Finelli:2008jv}.
This rotation leaves a distinct imprint on the CMB polarization by converting a portion of the parity-even $E$-modes into parity-odd $B$-modes, generating non-zero parity-odd $EB$ and $TB$ cross-power spectra which are expected to vanish in the absence of such a parity-violating effect \cite{zhaoFluctuationsCosmologicalBirefringence2014, zhaoDetectingRelicGravitational2014}. Given its lower cosmic variance compared to the $TB$ power spectrum, the $EB$ power spectrum is widely used to constrain the cosmic birefringence angle \cite{WMAP:2008lyn,QUaD:2008ado,WMAP:2010qai,planckcollaborationPlanckIntermediateResults2016b,Dou:2024wdy}.

Unfortunately, miscalibration of instrumental polarization angles of detectors would mimic a spurious rotation of the polarization plane, causing a degeneracy between the birefringence angle $\beta$ and the miscalibration angle $\alpha$ at each frequency band \cite{Keating:2012ge,Miller:2009pt}. The polarized Galactic foregrounds suffer from miscalibration but almost being immune to cosmic birefringence, enabling us to disentangle the physical $\beta$ from instrumental systematics. A novel technique \cite{minamiSimultaneousDeterminationCosmic2019,minamiSimultaneousDeterminationCosmic2020,minamiDeterminationMiscalibratedPolarization2020,minamiNewExtractionCosmic2020} (\revise{referred to as the ``Minami-Komatsu estimator''}) exploit the $EB$ cross-frequency power spectra (along with the prior knowledge of the intrinsic $EB$ correlation of Galactic foregrounds \cite{diego-palazuelosCosmicBirefringencePlanck2022}) to simultaneously determine the $\beta$ and $\alpha$ angles. The nearly full-sky analyses on \textit{WMAP} and \textit{Planck} data, based on this method, have reported tentative evidence for an isotropic CB signal of $\beta\simeq0.3^\circ$ at $2.4$ to $3.6\sigma$ significance \cite{diego-palazuelosCosmicBirefringencePlanck2022,eskiltFrequencyDependentConstraintsCosmic2022,eskiltImprovedConstraintsCosmic2022,eskiltCosmoglobeDR1Results2023}. Instead of running a full Markov Chain Monte Carlo (MCMC) sampler, Refs. \cite{diego-palazuelosCosmicBirefringencePlanck2022,diego-palazuelosRobustnessCosmicBirefringence2023,hozLiteBIRDScienceGoals2025} adopt a semi-analytical methodology with a small-angle approximation that significantly reduces computational time, using the Fisher matrix to derive the uncertainties. Besides, a field-level approach using hybrid internal linear combination (ILC) found a consistent angle with a $2.7\sigma$ detection from \textit{Planck} data \cite{remazeillesFieldlevelConstraintsCosmic2025}. A recent analysis of the Atacama Cosmology Telescope (ACT) Data release 6 (DR6) \cite{ACT:2025fju} resulted in $\beta=0.215^\circ\pm0.074^\circ$ rejecting $\beta=0$ with a significance of $2.9\sigma$ \cite{diego-palazuelosCosmicBirefringenceAtacama2025}.

Further CMB polarization data are required to give a tighter constraint on CB and a more favorable evidence for parity violation. The Ali CMB Polarization Telescope (AliCPT) is a ground-based CMB experiment located in Tibet, China, mainly focusing on the high-precision measurements of CMB polarization in the northern hemisphere at 95 GHz and 150 GHz \cite{liProbingPrimordialGravitational2019}. Currently in operation, this telescope is dedicated to the scientific goals of verifying the parity violation and detecting the primordial gravitational waves (PGWs) through the CMB $B$-mode polarization \cite{ghoshPerformanceForecastsPrimordial2022,Dou:2023vqg}. In this work, we apply a semi-analytical maximum-likelihood method \cite{diego-palazuelosRobustnessCosmicBirefringence2023} to the AliCPT and \textit{Planck} simulations, in order to derive the expected constrains on the isotropic CB angle for the AliCPT experiment.

This paper is organized as follows. In Sec.~\ref{sec:meth}, we describe the methodology used to simultaneously estimate the cosmic birefringence and miscalibration angles. We detail the simulations and masks in Sec.~\ref{sec:sims}, and report our results in Sec.~\ref{sec:res}. We discuss the results and make final conclusions in Sec.~\ref{sec:conclusion}.

\section{Methodology}
\label{sec:meth}

The methodology we employ in this paper is similar to the semi-analytical \revise{Minami-Komatsu} estimator presented in Refs. \cite{diego-palazuelosCosmicBirefringencePlanck2022,diego-palazuelosRobustnessCosmicBirefringence2023,hozLiteBIRDScienceGoals2025}, summarized as follows.

\subsection{Formula of the observed \textit{EB} spectrum}
\label{sec:form_eb_sp}

Miscalibration angles (of the $i$-th frequency band), $\alpha_i$, would rotate all the sky signals received by the detectors on the focal plane, including the CMB and polarized Galactic foreground emissions. If we assume that the pseudo-scalar field $\phi$ varies slowly in spacetime, the cosmic birefringence angle $\beta$ would be proportional to the propagation distance of photons in the field. Therefore, the birefringence effect on Galactic foregrounds is negligible compared to that on the CMB photons emitted from the last scattering surface. However, a positive dust $TB$ correlation was detected by \textit{Planck} \cite{planckcollaborationPlanck2018Results2020e}, which has been attributed to a misalignment between local magnetic fields
and dust filaments \cite{huffenbergerPowerSpectraPolarized2020, clarkOriginParityViolation2021}, and might lead to an intrinsic dust $EB$ correlation.

The basic principle of our method is exploiting the cross-frequency power spectra of the multi-channel sky maps containing CMB and foreground components to break the degeneracy between $\alpha_i$ and $\beta$. The $E$- and $B$-mode spherical harmonic coefficients at a frequency band $i$ are:
\begin{align}
    \begin{pmatrix}
    E_{\ell m}^{i, \rm o} \\
    B_{\ell m}^{i, \rm o}
    \end{pmatrix}
    &=
    \mathbf{R}(\alpha_i)
    \begin{pmatrix}
    E_{\ell m}^{i, \rm fg} \\
    B_{\ell m}^{i, \rm fg}
    \end{pmatrix}
    +
    \mathbf{R}(\alpha_i+\beta)
    \begin{pmatrix}
    E_{\ell m}^{i, \rm CMB} \\
    B_{\ell m}^{i, \rm CMB}
    \end{pmatrix}
    +
    \begin{pmatrix}
    E_{\ell m}^{i, \rm N} \\
    B_{\ell m}^{i, \rm N}
    \end{pmatrix}
    \,,
    \\
    &\text{with}\quad
    \mathbf{R}(\theta)=
    \begin{pmatrix}
    c_{2\theta} & -s_{2\theta}\\
    s_{2\theta} & c_{2\theta}
    \end{pmatrix}
    \,.
\end{align}
where throughout this work, the superscripts ``o/fg/CMB/N'' stand for the observed sky, the intrinsic\footnote{Here, by ``intrinsic'' we mean the foreground and CMB components are before undergoing the CB effect.} Galactic foreground, intrinsic CMB, and noise components, respectively; hereafter ``$c_x$/$s_x$/$t_x$'' denote the cosine, sine, and tangent of an angle $x$, respectively. Note that unless otherwise specified, all the CMB and foreground spherical harmonic coefficients and angular power spectra are convolved with the beam transfer and pixel window functions $b_\ell^i$ and $p_\ell^i$.

Assuming no intrinsic $EB$ correlations in the CMB or noise, we can derive the $EB$ power spectrum for a single channel $i$ (see Eq.~(9) of \cite{minamiSimultaneousDeterminationCosmic2019}):
\begin{equation}\label{eq:clEB_1}
    C_\ell^{EB,i,\rm o}=\frac{t_{4\alpha_i}}{2}\left({C}_\ell^{EE,i,\rm o}-{C}_\ell^{BB,i,\rm o}\right)+\frac{s_{4\beta}}{2c_{4\alpha_i}}\left({C}_\ell^{EE,i,\rm CMB}-{C}_\ell^{BB,i,\rm CMB}\right)+\frac{C_\ell^{EB,i,\rm fg}}{c_{4\alpha_i}}\,,
\end{equation}
and the cross-frequency $EB$ correlation between the $i$-th and $j$-th frequency bands:
\begin{equation}\label{eq:clEB_2}
    \begin{aligned}
    C_\ell^{E_iB_j,\rm o}=\ &\frac{1}{c_{4\alpha_i}+c_{4\alpha_j}}\bigg(s_{4\alpha_j}C_\ell^{E_iE_j,\rm o}-s_{4\alpha_i}C_\ell^{B_iB_j,\rm o}
    +2c_{2\alpha_i}c_{2\alpha_j}C_\ell^{E_iB_j,\rm fg}+2s_{2\alpha_i}s_{2\alpha_j}C_\ell^{B_iE_j,\rm fg}\bigg)
    \\
    &+\frac{s_{4\beta}}{2c_{2\alpha_i+2\alpha_j}}\left(C_\ell^{E_iE_j,\rm CMB}-C_\ell^{B_iB_j,\rm CMB}\right)\,.
    \end{aligned}
\end{equation}
For $i=j$, Eq.~\eqref{eq:clEB_2} reduces to Eq.~\eqref{eq:clEB_1}. Eq.~\eqref{eq:clEB_1} assumes that the auto spectra of the $E$- and $B$-mode noise cancel out in average: $\langle C_\ell^{EE,i,\rm N}-C_\ell^{BB,i,\rm N}\rangle=0$.

Auto-frequency spectra ($i=j$) are excluded from our analysis to avoid the \revise{noise bias and correlated systematics (including the temperature-to-polarization leakage, beam imperfection, and cross-polarization effects, etc.) in the observed auto-power spectra. The noise-dominated $EE$ and $BB$ spectra might introduce numerical instabilities in the computation of maximum-likelihood solutions. Despite the high signal-to-noise ratio (S/N) of $E$-modes for AliCPT, its measurement of CMB $B$-modes is still noise-dominated \cite{chen01,Zhang:2024kmp}. Hence, we divide each AliCPT band into two independent data splits (95A, 95B, 150A, 150B) like \textit{Planck}, and we validate the use of them in App.~\ref{app:valid_data_split}.}

Under the small-angle approximation\footnote{This approximation is guaranteed by the current constraints on the birefringence angle (e.g., \cite{eskiltImprovedConstraintsCosmic2022}) and the strict control over the instrumental errors in the future experiments (e.g., \cite{Vielva_2022,ritaccoAbsoluteReferenceMicrowave2024}).} ($|\alpha_i|\lesssim5^\circ$ and $|\beta|\lesssim5^\circ$), Eq.~\eqref{eq:clEB_2} can be simplified as:
\begin{equation}\label{eq:clEB_3}
    C_\ell^{E_iB_j,\rm o}\approx 2\alpha_jC_\ell^{E_iE_j,\rm o}-2\alpha_iC_\ell^{B_iB_j,\rm o}+2\beta\left(C_\ell^{E_iE_j,\rm CMB}-C_\ell^{B_iB_j,\rm CMB}\right)+C_\ell^{E_iB_j,\rm fg}\,.
\end{equation}
Eqs.~\eqref{eq:clEB_2} and \eqref{eq:clEB_3} are commonly used to construct a maximum likelihood estimator to simultaneously measure $\alpha_i$ and $\beta$ \cite{minamiNewExtractionCosmic2020,eskiltImprovedConstraintsCosmic2022,diego-palazuelosCosmicBirefringencePlanck2022}. Besides the observed power spectra that could be directly computed from sky maps, the equations require the prior knowledge of the foreground $EB$ spectrum and the CMB $EE$ and $BB$ power spectra. We adopt the best-fit $\rm \Lambda$CDM cosmological parameters to compute the CMB power spectra using \texttt{CAMB}\footnote{\url{https://github.com/cmbant/CAMB}}  \cite{lewisEfficientComputationCMB2000} package (see Sec.~\ref{sec:sims}). The contribution from the tensor $B$ modes constrained by \textit{BICEP/Keck} observations \cite{collaborationBICEPKeckXIII2021} is negligible compared to the CMB $E$-mode power spectrum.

\subsection{Modeling the dust \textit{EB} correlation}

We conservatively consider a non-zero foreground $EB$ correlation even though the current CMB measurements find it to be statistically consistent with zero \cite{martireCharacterizationPolarizedSynchrotron2022,planckcollaborationPlanck2018Results2020e}. Two independent approaches of modeling the dust $EB$ spectrum has yielded consistent results for \textit{Planck} Public Release 4 (PR4) data \cite{Planck:2020olo} in \cite{diego-palazuelosCosmicBirefringencePlanck2022}: one predicts the dust $EB$ correlation based on a filament misalignment model \cite{huffenbergerPowerSpectraPolarized2020,clarkOriginParityViolation2021}, which is the only physical foreground $EB$ model available today (e.g., \cite{eskiltFrequencyDependentConstraintsCosmic2022,eskiltImprovedConstraintsCosmic2022}); and the other utilizes foreground templates given by a component separation technique, e.g., \texttt{Commander} \cite{eriksenJointBayesianComponent2008} to compute the foreground $EB$ spectra (e.g., \cite{diego-palazuelosRobustnessCosmicBirefringence2023,hozLiteBIRDScienceGoals2025}).

The first method assumes that the dust $EB$ spectrum is proportional to the observed dust $TB$, i.e., $C_\ell^{EB, \rm dust}/C_\ell^{EE, \rm dust}\propto C_\ell^{TB, \rm dust}/C_\ell^{TE, \rm dust}$, where the $TB$ and $TE$ spectra are computed from \textit{Planck} 353 GHz observations \cite{eskiltImprovedConstraintsCosmic2022}. \revise{Refs. \cite{Hervias-Caimapo:2021zue,Hervias-Caimapo:2024ili} calibrate the filamentary dust model to \textit{Planck} and HI observations to reproduce parity-violating foreground spectra, and find that the results of measuring CB using the calibrated model as a prior are consistent with \cite{diego-palazuelosCosmicBirefringencePlanck2022}. However, the \textit{Planck}'s sensitivity to $EB$ correlations is insufficient to test the filament model with high significance, and tighter constraints on the misalignment angle from future high-frequency observations are required to improve predictions of the dust $EB$ spectrum. The second approach is better suited for simulation analysis, as it allows quantification of the estimator's intrinsic error due to approximations under perfect foreground modeling.} Nevertheless, it should be treated with caution in the real case since the \texttt{Commander} foreground templates are not signal-dominated in terms of $EB$ correlations \cite{diego-palazuelosRobustnessCosmicBirefringence2023,Planck:2018yye}.

In our analysis of simulations, we draw on the latter approach, taking the fiducial foreground maps based on the \texttt{PySM}\footnote{\url{https://github.com/galsci/pysm/tree/main}} models \cite{thornePythonSkyModel2017,zoncaPythonSkyModel2021,groupFullskyModelsGalactic2025} (see Sec.~\ref{sec:sims}) as our foreground templates.
We compute $C_\ell^{EB, \rm fg}$ from the foreground templates which are exactly the input foreground maps in our simulations. Finally, we take $\mathcal{A}C_b^{EB,\rm fg}$ as the foreground $EB$ correlation in our model, with the $EB$ spectrum of templates modified by a free amplitude parameter $\mathcal{A}$ to be fit along with $\alpha_i$ and $\beta$.

\subsection{Construction of the likelihood and the covariance matrix}

From Eq.~\eqref{eq:clEB_3}, we could build a Gaussian likelihood:
\begin{equation}\label{eq:llh}
\begin{aligned}
    -2\ln\mathcal{L}\supset\sum_{i,j,p,q}\sum_{b}\bigg[&C_b^{E_iB_j,\rm o}-2\alpha_jC_b^{E_iE_j,\rm o}+2\alpha_iC_b^{B_iB_j,\rm o}-2\beta\left(C_b^{E_iE_j,\rm CMB}-C_b^{B_iB_j,\rm CMB}\right)
    \\&-\mathcal{A}C_b^{E_iB_j,\rm fg}\bigg]
    \mathbf{C}^{-1}_{ijpqb}\bigg[C_b^{E_pB_q,\rm o}-2\alpha_qC_b^{E_pE_q,\rm o}+2\alpha_pC_b^{B_pB_q,\rm o}
    \\&-2\beta\left(C_b^{E_pE_q,\rm CMB}-C_b^{B_pB_q,\rm CMB}\right)-\mathcal{A}C_b^{E_pB_q,\rm fg}\bigg]\,.
\end{aligned}
\end{equation}
Here the $\supset$ symbol indicates that the likelihood form is incomplete, since the $\ln|\mathbf{C}|$ term is not included. This log-determinant is automatically considered in the iterative process, as shown in App.~B of \cite{diego-palazuelosRobustnessCosmicBirefringence2023} and discussed later. In Eq.~\eqref{eq:llh} we sum over all combinations of two frequency channel pairs $(i,j)$ and $(p,q)$ with $i\neq j$ and $p\neq q$, and all the power spectra and the covariance matrix are binned with a spacing of $\Delta\ell=20$ at $\ell\in[51, 1490]$\footnote{We validate the choice of the multipole range for AliCPT in App.~\ref{app:vary-lrange}.} adopted by the \textit{Planck} analysis \cite{minamiNewExtractionCosmic2020,diego-palazuelosCosmicBirefringencePlanck2022}:
\begin{equation}
    C_b^X=\frac{1}{\Delta\ell}\sum_{\ell\in b}C_\ell^X\,,\quad \mathbf{C}_b=\frac{1}{\Delta\ell^2}\sum_{\ell\in b}\mathbf{C}_\ell\,.
\end{equation}
In this work, we compute the pseudo-$C_\ell$s on the masked sky and decouple them using \texttt{NaMASTER}\footnote{\url{https://github.com/LSSTDESC/NaMaster}} \cite{alonsoUnifiedPseudoC_2019} without $E/B$ purification. Although the standard pseudo-$C_\ell$ method is slightly sub-optimal for the variance of the $B$-mode power spectrum due to $E$-to-$B$ leakage \cite{zhaoSeparatingTypesPolarization2010,alonsoUnifiedPseudoC_2019,ghoshEndingPartialSky2021}, we have verified that the $B$-mode uncertainty related to the algorithm is still negligible compared to the lensing $B$-mode power spectrum.

The full $N_\nu^2 N_b\times N_\nu^2 N_b$ covariance matrix is defined as (where $N_\nu$ is the number of frequency channels and $N_b=72$ is the number of multipole bins):
\begin{equation}\label{eq:cov_def}
\begin{aligned}
    \mathbf{C}_{ijpqbb'}
    \\
    ={\rm Cov}\bigg[&C_b^{E_iB_j,\rm o}-2\alpha_jC_b^{E_iE_j,\rm o}+2\alpha_iC_b^{B_iB_j,\rm o}-2\beta\left(C_b^{E_iE_j,\rm CMB}-C_b^{B_iB_j,\rm CMB}\right)-\mathcal{A}C_b^{E_iB_j,\rm fg}, 
    \\& C_{b'}^{E_pB_q,\rm o}-2\alpha_qC_{b'}^{E_pE_q,\rm o}+2\alpha_pC_{b'}^{B_pB_q,\rm o}-2\beta\left(C_{b'}^{E_pE_q,\rm CMB}-C_{b'}^{B_pB_q,\rm CMB}\right)-\mathcal{A}C_{b'}^{E_pB_q,\rm fg}\bigg]\,.
\end{aligned}
\end{equation}
In the absence of a reliable model of Galactic foreground spectra, we can approximate the covariance assuming that the spherical harmonic coefficients are Gaussian as:
\begin{equation}\label{eq:cov_XYZW}
    {\rm Cov}\left[C_b^{XY}, C_{b'}^{ZW}\right]\approx \frac{1}{(2\ell_b+1)f_{\rm sky}\Delta\ell}\delta_{bb'}\left[C_b^{XZ}C_b^{YW}+C_b^{XW}C_b^{YZ}\right]\,,
\end{equation}
where $\ell_b$ is the center multipole of the bin $b$. The effective sky fraction of an apodized mask is given by $f_{\rm sky}=\langle W^2\rangle^2/\langle W^4\rangle$ where $W$ is the (nonbinary) weight of the mask. Here we neglect the $b$-$b'$ coupling (i.e., off-diagonal elements) in the binned covariance matrix since the $\ell$-$\ell'$ correlations are reduced by binning and mask apodization. This approximation is implied through reducing $\mathbf{C}_{ijpqbb'}$ to $\mathbf{C}_{ijpqb}$ and $\sum_{bb'}$ to $\sum_{b}$ in Eq.~\eqref{eq:llh}.

As previously mentioned, the CMB $EE$ and $BB$ power spectra are calculated from the best-fit $\rm \Lambda$CDM model and are smoothed with the beam and pixel window functions. A model that we are convinced of should not alter across realizations in a certain ensemble, and therefore the covariance between the CMB spectrum and itself or other spectra should not appear in the full covariance matrix. Similarly, if we model the foreground $EB$ spectrum by a dust filamentary model, the covariance matrix would have no contribution from the foreground spectrum either, and thus it could be derived from only the observed spectra using Eq.~\eqref{eq:cov_XYZW}, for example in Eq.~(12) of \cite{minamiSimultaneousDeterminationCosmic2020}.

Otherwise, if we build a foreground template and regard it as a random realization of some underlying foreground power spectra, the uncertainties of its power spectra should also contribute to the total covariance. In summary, the covariance defined in Eq.~\eqref{eq:cov_def} will contain three terms (note the minus sign before the last term):
\begin{equation}\label{eq:cov_full}
    \mathbf{C}_{ijpqb}=\frac{1}{(2\ell_b+1)f_{\rm sky}\Delta\ell}\left[\mathbf{C}^{\rm o*o}_{ijpqb}+\mathbf{C}^{\rm fg*fg}_{ijpqb}-\mathbf{C}^{\rm fg*o}_{ijpqb}\right]\,,
\end{equation}
including the covariance of the observed spectra, the covariance of the foreground spectra, and the cross covariance between the observed and foreground spectra, with the full expression of each term presented in App.~\ref{app:der-cov} (also see App.~A of \cite{hozLiteBIRDScienceGoals2025}). We note that $\mathbf{C}^{\rm fg*fg}_{ijpqb}-\mathbf{C}^{\rm fg*o}_{ijpqb}$ is always negative, reducing the total covariance and leading to a lower uncertainty compared to the case when ignoring the foreground $EB$ correlation ($\mathcal{A}=0$). This is because $\mathbf{C}^{\rm fg*o}_{ijpqb}\approx2\mathbf{C}^{\rm fg*fg}_{ijpqb}$ when $\mathcal{A}\approx1$, given the same fluctuations between the foreground template and the observed sky map (see App.~D of \cite{diego-palazuelosRobustnessCosmicBirefringence2023}).

\subsection{Iterative algorithm for the maximum likelihood solution}

To minimize the minus-log-likelihood in Eq.~\eqref{eq:llh}, we adopt an iterative algorithm to determine the maximum likelihood solution. This algorithm significantly accelerates computation while maintaining the accuracy and precision of the full-likelihood MCMC sampling.

We start with an initial value of free parameters $\mathbf{x}=(\mathcal{A}, \beta, \alpha_i)=(1,0,0)$ and implement several iterations updating them until convergence. Each iteration contains two steps: first compute the covariance matrix with the parameters $\mathbf{x}=(\mathcal{A}, \beta, \alpha_i)$ assumed to be known and fixed; then differentiate Eq.~\eqref{eq:llh} with respect to each parameter and solve $\partial(-2\ln\mathcal{L})/\partial\mathbf{x}_i=0$ to derive a new estimate of $\mathbf{x}$. Through this process we calculate the maximum likelihood solution analytically under a fixed covariance matrix, and therefore the $\ln|\mathbf{C}|$ term in Eq.~\eqref{eq:llh} needs not to be explicitly considered. After the iteration we update both the parameters $\mathbf{x}$ and the covariance matrix $\mathbf{C}$, and calculate next best-fit solution until $\mathbf{x}$ converges. The number of iterations is set to be 10 in our pipeline, since we have verified that all the parameters converge after several iterations.

The linear equations $\partial(-2\ln\mathcal{L})/\partial\mathbf{x}_n=0$ with fixed $\mathbf{C}$ is equivalent to a linear system $\sum_n\mathbf{A}_{mn}\mathbf{x}_n=\mathbf{b}_m$. See App.~\ref{app:lin-eqs} for the complete form of $\mathbf{A}_{mn}$ and $\mathbf{b}_m$ (also see App.~A of \cite{hozLiteBIRDScienceGoals2025}). In addition, the covariance matrix of fitted parameters can be approximated as the inverse of the Fisher matrix: $[{\rm Cov}(\mathbf{x})]^{-1}_{mn}=-\frac{\partial^2\ln\mathcal{L}}{\partial\mathbf{x}_m\partial\mathbf{x}_n}=\mathbf{A}_{mn}$. Hence we can calculate the theoretical uncertainty of parameters from the linear system: $\sigma^2(\mathbf{x}_n)=[{\rm Cov}(\mathbf{x})]_{nn}=[\mathbf{A}^{-1}]_{nn}$.

\section{Simulations and masks}
\label{sec:sims}

\begin{figure}[tbp]
    \centering
    \begin{subfigure}[b]{0.49\textwidth}
        \centering
        \includegraphics[width=\textwidth]{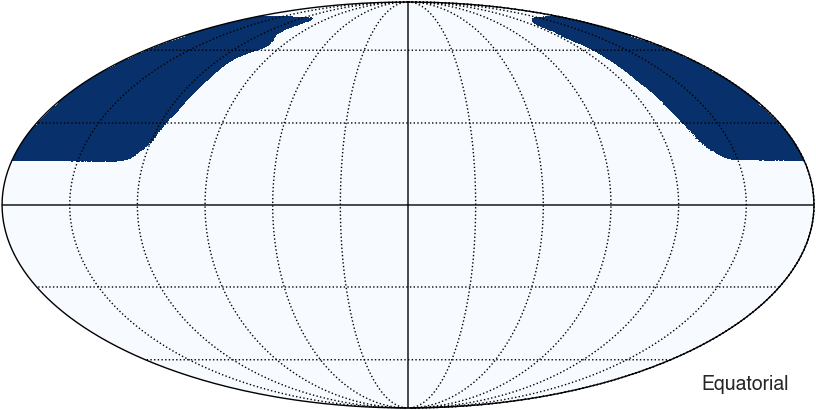}
    \end{subfigure}
    \begin{subfigure}[b]{0.49\textwidth}
        \centering
        \includegraphics[width=\textwidth]{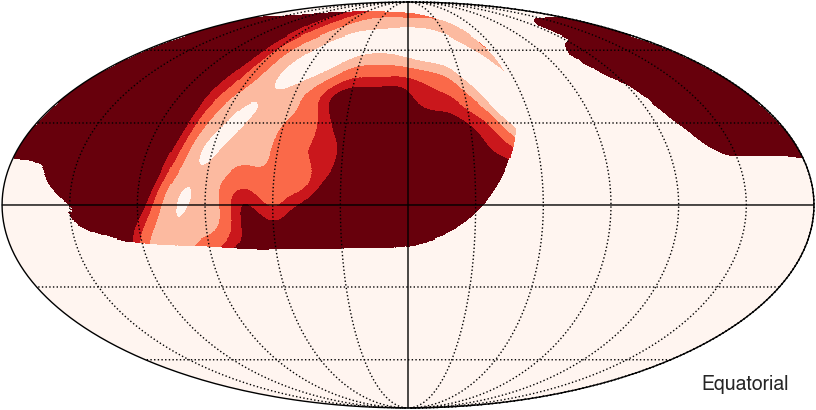}
    \end{subfigure}
    \caption{Masks in Equatorial coordinates. \textit{Left}: the D30$\mu$K mask adopted for the deep-scan patch, with a sky fraction of 11\%. \textit{Right}: the W90, W80, W70, and W60 masks used for analyses of the wide survey region, indicated with colors from lightest to darkest red, with $f_{\rm sky}=38\%$, 32\%, 28\%, and 25\%, respectively.}
    \label{fig:msk}
\end{figure}

The AliCPT experiment plans to implement two scanning strategies \cite{liProbingPrimordialGravitational2019,Dou:2024spy}: the ``deep-scan'' strategy targeting the lowest-foreground-contaminated region in the northern hemisphere with a sky coverage of $17\%$, mainly focusing on the detection of primordial $B$ modes; and the ``wide-scan'' strategy covering most of the northern sky with a sky fraction of about 40\% but a higher noise level compared to the deep scan.

The deep survey covers a trapezoid-shaped sky patch centering at ${\rm RA}=170^\circ$ and ${\rm DEC}=40^\circ$ in Celestial coordinates. We use a mask that preserves the pixels with noise standard deviation lower than 30 $\mu$K-pixel at the 150 GHz channel, whose sky fraction is about 11\% (see the left plot of Fig.~\ref{fig:msk}), referred to as ``D30$\mu$K'' mask in the following. 

For the wide survey scenario, we apply the \textit{Planck} Galactic masks\footnote{\url{http://pla.esac.esa.int/pla/aio/product-action?MAP.MAP_ID=HFI_Mask_GalPlane-apo0_2048_R2.00.fits}} with different sky coverages from 60\% to 90\% to the wide survey region. \revise{Similarly, We remove the pixels with polarized noise standard deviation higher than 30 $\mu$K at 150 GHz.} The sky fractions of the obtained masks are 38\%, 32\%, 28\%, and 25\%, corresponding to the 90\%, 80\%, 70\%, and 60\% \textit{Planck} Galactic masks, respectively. We refer to them as ``W90, W80, W70, and W60'' masks, as shown in the right plot of Fig.~\ref{fig:msk}. All the masks are apodized with C2 type\footnote{\url{https://namaster.readthedocs.io/en/latest/source/sample_masks.html}} and $1^\circ$ FWHM in the computation of power spectra.

\begin{table}[tbp]
    \centering
    \caption{Instrumental characteristics of the six AliCPT and \textit{Planck} HFI frequency channels. $\sigma_n^{\rm P}$ represents the noise level for polarized full-mission maps, where the values in parentheses are the ones for the wide-scan strategy (D30$\mu$K) of AliCPT channels, otherwise for the deep-scan scenario (W90).}
    \begin{tabular}{c|cc|cccc}
        \hline
        \hline
        Experiment&\multicolumn{2}{c|}{AliCPT}&\multicolumn{4}{c}{PR4 HFI} \\
        \hline
        Frequency (GHz)&95&150&100&143&217&353 \\
        \hline
        FWHM (arcmin)&19&11&9.7&7.3&5.0&4.9 \\
        \hline
        $\sigma_n^{\rm P}$ ($\mu$K-arcmin)&18 (34)&24 (44)&81&66&92&399\\
        \hline
        \hline
    \end{tabular}
    \label{tab:instr}
\end{table}

\begin{table}[tbp]
    \centering
    \caption{Four simulation scenarios used in this study, with different input birefringence angles $\beta$, miscalibration angles $\alpha_i$, and sky models employed to produce foreground simulations using \texttt{PySM3}.}
    \begin{tabular}{c|c|c|c}
        \hline
        \hline
        Dataset & $\beta$ [deg] & $\alpha_i$ [deg] & Foreground Complexity \\
        \hline
        1 & 0 & 0 & Low (d9,s4,f1,a1,co1) \\
        2 & 0 & 0 & High (d12,s7,f1,a2,co3) \\
        3 & 0.3 & $U(-0.3,0.3)$ & Low \\
        4 & 0.3 & $U(-0.3,0.3)$ & High \\
        \hline
        \hline
    \end{tabular}
    \label{tab:input-rot}
\end{table}

In this work, we consider the data combination of the one-year AliCPT and the \textit{Planck} Public Release 4 \cite{Planck:2020olo} High Frequency Instrument (HFI) observations, with the instrumental parameters summarized in Tab.~\ref{tab:instr}. We generate four mock datasets (named Dataset 1-4) with different foreground complexities and input rotation angles described in Tab.~\ref{tab:input-rot}. For each dataset, we produce 100 polarized sky maps across six frequency bands at \texttt{HEALPix}\footnote{\url{https://healpix.jpl.nasa.gov}} \cite{gorskiHEALPixFrameworkHigh2005} resolution $N_{\rm side}=1024$.

\begin{figure}[tbp]
    \centering
    \begin{subfigure}[b]{0.9\textwidth}
        \centering
        \includegraphics[width=\textwidth]{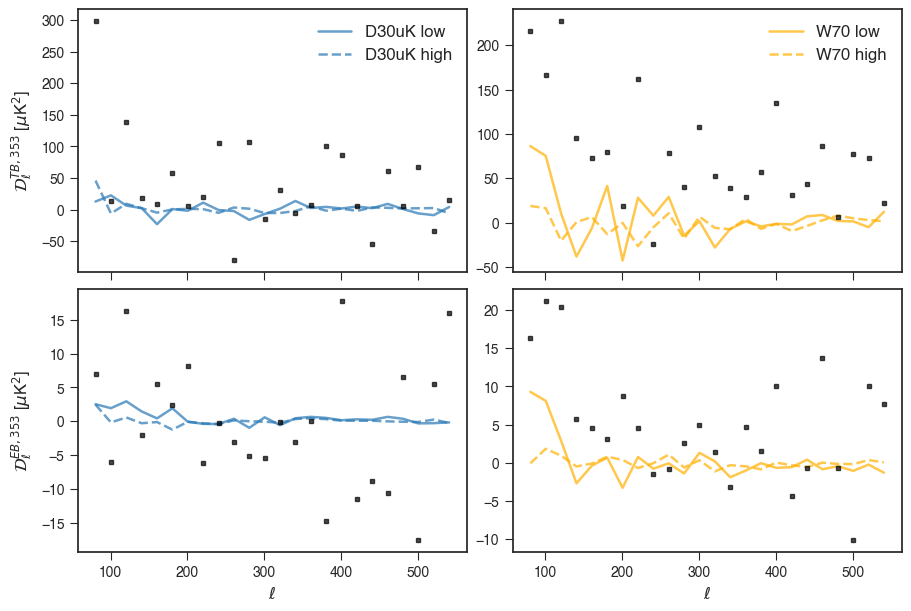}
    \end{subfigure}
    \caption{$TB$ (upper) and $EB$ (lower) power spectra of low-complexity (solid) and high-complexity (dashed) foreground templates at 353 GHz for the two survey regions, D30$\mu$K (left blue) and W70 (right yellow). The black squares represent power spectra of the PR4 353 GHz map applied with the two masks respectively.}
    \label{fig:cl-353fg}
\end{figure}

We simulate the Galactic foregrounds including thermal dust, synchrotron, free-free, AME, and CO line emissions, based on the \texttt{PySM} \cite{thornePythonSkyModel2017,zoncaPythonSkyModel2021,groupFullskyModelsGalactic2025} models. We adopt the low- and high-complexity fiducial sky scenarios described on the PanEx website\footnote{\url{https://galsci.github.io/blog/2022/common-fiducial-sky}}. Specifically, the \texttt{PySM3} \texttt{d9,s4,f1,a1,co1} emissions make up the low-complexity sky model, with the \textit{Planck} GNILC thermal dust template scaled by a modified black body spectrum, the \textit{WMAP} 9-year synchrotron template scaled by a power law SED using fixed spectral indices\footnote{To be specific, $\beta_d=1.48$, $T_d=\SI{19.6}{K}$ for dust and $\beta_s=-3.1$ for synchrotron.}, and unpolarized free-free, Anomalous Microwave Emission (AME), and CO line emissions. The high-complexity model consists of the \texttt{PySM3} \texttt{d12,s7,f1,a2,co3} emissions, with spatially varying SEDs for dust and synchrotron, and faint polarization for AME and CO. The high-complexity foreground includes more small scales and frequency decorrelation effects than the low-complexity one.

The thermal dust emission dominates the foreground contamination in the frequency bands considered in this work. We compare the dust $TB$ and $EB$ correlations of the two foreground models and the \textit{Planck} observations\footnote{The PR4 353 GHz map can be downloaded from \url{http://pla.esac.esa.int/pla/aio/product-action?MAP.MAP_ID=HFI_SkyMap_353-BPassCorrected-field-IQU_2048_R4.00_full.fits}.} by plotting their power spectra at 353 GHz in Fig.~\ref{fig:cl-353fg}. Here we apply an extra \textit{Planck} point source and CO mask\footnote{\revise{Here we use the f92 mask introduced in Sec.~\ref{sec:test-hfi}, available on the google drive link in \url{https://github.com/LilleJohs/Cosmic_Birefringence}.}} on both deep-scan and wide-scan regions when computing the power spectra. We notice that the $TB$ and $EB$ spectra of both low- (solid) and high-complexity (dashed) foreground models are consistent with zero for the D30$\mu$K region (left blue), while the low-complexity foreground spectra are larger than the high-complexity foreground spectra at large scales for the W70 region (right yellow). The PR4 353 GHz map exhibits a clear positive $TB$ correlation (black square), underscored by the \textit{Planck} paper \cite{planckcollaborationPlanck2018Results2020e}, which is not fully reproduced by our foreground models. 

The CMB maps are random realizations generated from the power spectra derived from the $\rm \Lambda$CDM model with gravitational lensing but no tensor perturbations, computed from the \textit{Planck} 2018 best-fit cosmological parameters\footnote{The six $\rm \Lambda$CDM parameters are: dark matter density $\Omega_ch^2=0.120$, baryon density $\Omega_bh^2=0.02237$, scalar spectral index $n_s=0.9649$, optical depth $\tau=0.0544$, Hubble constant $H_0=67.36\ {\rm km\ s^{-1}\ Mpc^{-1}}$ and the primordial comoving curvature power spectrum amplitude $A_s=2.10\times10^{-9}$.} \cite{aghanimPlanck2018Results2020a} using the \texttt{CAMB} \cite{lewisEfficientComputationCMB2000} package. For Dataset 1 and 2, we do no rotation to the sky maps, i.e., $\alpha_i=\beta=0$. For Dataset 3 and 4, we rotate the CMB by $\alpha_i+\beta$ and the foreground maps by only $\alpha_i$, with $\beta=0.3^\circ$ and miscalibration angles drawn from a uniform distribution of $[-0.3^\circ,0.3^\circ]$, \revise{comparable to the $2\sigma$ (95\%) CI. of \textit{Planck}'s miscalibration angles (see Tab.~1 of \cite{diego-palazuelosCosmicBirefringencePlanck2022})}. After rotation, the CMB and foreground simulations are coadded and convolved with the nominal Gaussian beams listed in Tab.~\ref{tab:instr}. The Gaussian white noise for AliCPT bands and the PR4 NPIPE noise simulations for \textit{Planck} HFI bands are then added to the sky signal maps at their corresponding channels. The NPIPE A/B noise simulations are downloaded from the \textit{Planck} Legacy Archive (PLA)\footnote{\url{https://pla.esac.esa.int}}. To avoid noise bias and correlated systematics, we produce A/B data splits for each sky simulation, adding two independent noise simulations to the same sky signal map. The noise level of splits is $\sqrt{2}$ larger than that of the full map. We compute $EE$, $BB$ and $EB$ power spectra between all data split pairs, e.g., 95A$\times$95B.

\section{Results}
\label{sec:res}

\subsection{Test on Planck HFI simulations}
\label{sec:test-hfi}

First we test our methodology on the PR4 HFI simulations without including AliCPT bands. To validate the pipeline against PR4 data results, we use two masks adopted in previous analyses of \textit{Planck} data \cite{minamiNewExtractionCosmic2020,diego-palazuelosCosmicBirefringencePlanck2022,eskiltImprovedConstraintsCosmic2022}, with the sky fraction of 92\% and 62\%, referred to as ``f92 and f62'' masks. Both masks exclude pixels where the CO line or point sources are bright, and the f62 mask further removes 30\% of the sky near the Galactic plane.

\begin{figure}[tbp]
    \centering
    \begin{subfigure}[b]{0.49\textwidth}
        \centering
        \includegraphics[width=\textwidth]{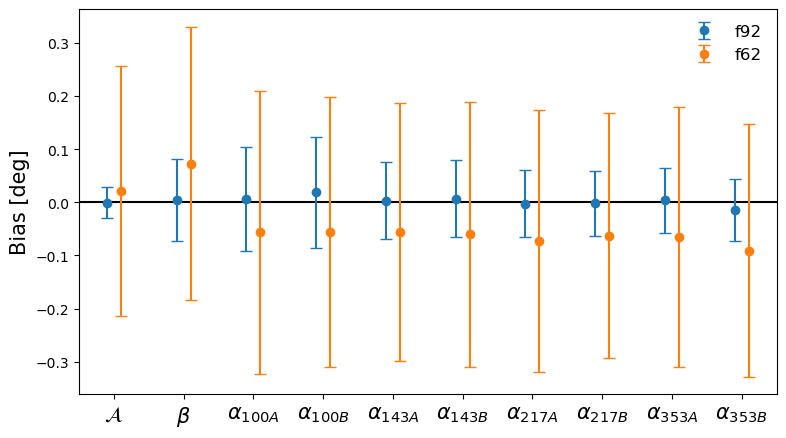}
    \end{subfigure}
    \begin{subfigure}[b]{0.49\textwidth}
        \centering
        \includegraphics[width=\textwidth]{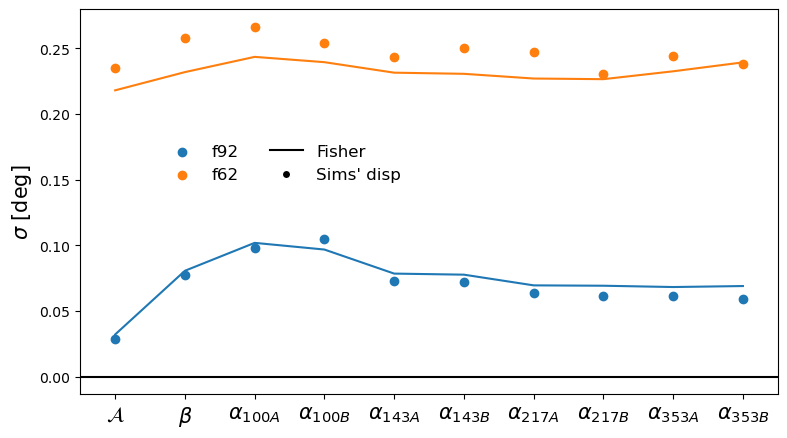}
    \end{subfigure}
    \caption{\textit{Left}: Bias of the measured parameters for HFI-only Dataset 1 simulations (low-FG, $\beta=\alpha_i=0$), with the 68\% C.L. of the simulations' dispersion taken as the error bar. \textit{Right}: Uncertainties of fitted parameters. The circles represent the dispersion of simulations, and the curves show the uncertainties predicted by the Fisher matrix.}
    \label{fig:hfi-low-null}
\end{figure}

\begin{table}[tbp]
    \centering
    \caption{Measured $\beta$ for HFI-only simulations across four datasets.}
    \begin{tabular}{c|c|c}
        \hline
        \hline
        Dataset & f92 & f62 \\
        \hline
        1 (low-FG, $\beta=\alpha_i=0$) & $0.00^\circ\pm0.08^\circ$ & $0.07^\circ\pm0.26^\circ$ \\
        2 (high-FG, $\beta=\alpha_i=0$) & $0.01^\circ\pm0.08^\circ$ & $0.06^\circ\pm0.28^\circ$ \\
        3 (low-FG, $\beta=0.3^\circ$, $\alpha_i\neq0$) & $0.30^\circ\pm0.08^\circ$ & $0.35^\circ\pm0.28^\circ$ \\
        4 (high-FG, $\beta=0.3^\circ$, $\alpha_i\neq0$) & $0.30^\circ\pm0.08^\circ$ & $0.34^\circ\pm0.28^\circ$ \\
        \hline
        \hline
    \end{tabular}
    \label{tab:hfi-beta}
\end{table}

For Dataset 1 (low-complexity foreground and no input rotation angles), the left panel of Fig.~\ref{fig:hfi-low-null} presents the bias (measured values minus input values) of 10 fitted parameters including the foreground amplitude $\mathcal{A}$, rotation angles $\beta$ and $\alpha_i$ for A/B splits of the four HFI bands. The circles in the right panel represent the $1\sigma$ uncertainties from the simulations' dispersion of measured parameters, while the curves represent the uncertainties calculated from the Fisher matrix. We see the two uncertainties are compatible, justifying the approximation in the computation of the covariance matrix. \revise{As expected, the Fisher formalism tends to underestimate the uncertainty of parameters for the f62 mask (orange) since the variance of the foreground emission and the cross covariance between the foreground and observed spectra are taken into accounted in the covariance matrix (remember the sum of these two terms is negative), but only one foreground realization was used in each dataset of simulations \cite{diego-palazuelosRobustnessCosmicBirefringence2023}.} Tab.~\ref{tab:hfi-beta} lists the uncertainties of measured $\beta$ (dispersion of simulations) across four simulation scenarios and two masks. We obtain \revise{$\hat{\beta}=0.00^\circ\pm0.08^\circ$ and  $\hat{\beta}=0.07^\circ\pm0.26^\circ$} for Dataset 1, $f_{\rm sky}=92\%$ and $f_{\rm sky}=62\%$ cases, respectively.


As shown in Tab.~\ref{tab:hfi-beta}, our estimator accurately estimates $\hat{\beta}\approx0.3^\circ$ for Dataset 3 and 4 simulations, with the uncertainties $\sigma(\beta)=0.08^\circ$ and $0.28^\circ$ for f92 and f62 cases, respectively. The uncertainties are consistent with those of previous PR4 works, e.g., Tab.~1 of \cite{diego-palazuelosCosmicBirefringencePlanck2022}.

\subsection{Results on AliCPT and HFI simulations}

\begin{figure}[tbp]
    \centering
    \begin{subfigure}[b]{0.9\textwidth}
        \centering
        \includegraphics[width=\textwidth]{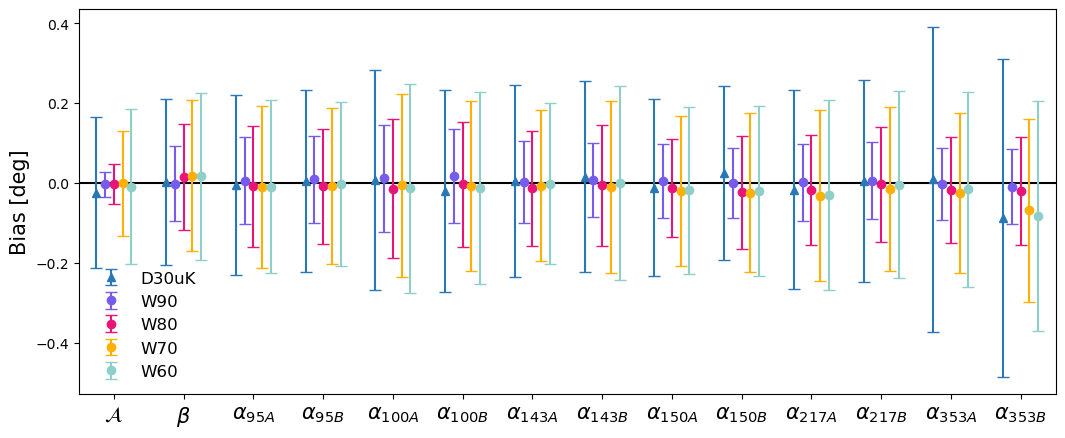}
    \end{subfigure}
    \caption{Bias of the measured parameters for AliCPT+HFI Dataset 1 (low-FG, $\beta=\alpha_i=0$) simulations.}
    \label{fig:ali-low-null}
\end{figure}

\begin{figure}[tbp]
    \centering
    \begin{subfigure}[b]{0.9\textwidth}
        \centering
        \includegraphics[width=\textwidth]{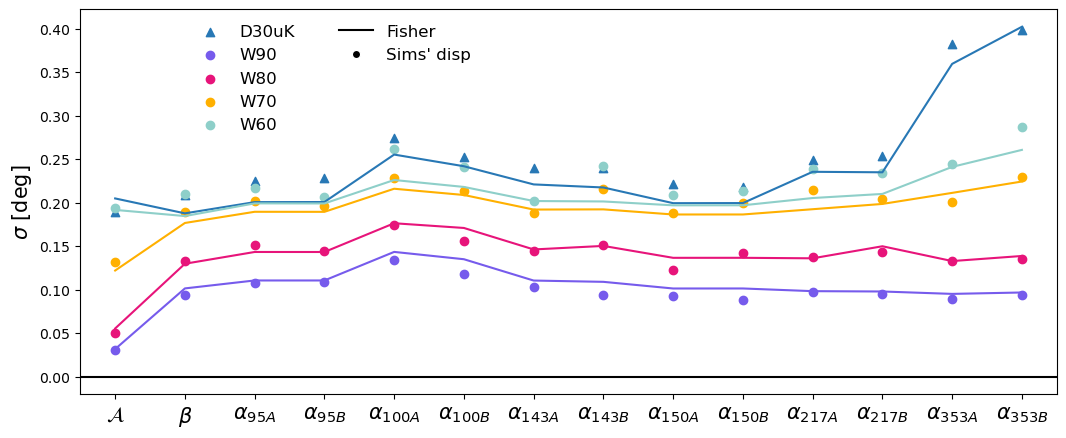}
    \end{subfigure}
    \caption{Uncertainties of the measured parameters for AliCPT+HFI Dataset 1 (low-FG, $\beta=\alpha_i=0$). The circles and triangles represent the dispersion of simulations, and the curves denote the uncertainties computed from the Fisher matrix.}
    \label{fig:sigma-ali-low-null}
\end{figure}

\begin{table}[tbp]
    \centering
    \caption{Measured $\beta$ for AliCPT and PR4 HFI simulations across four datasets.}
    \begin{tabular}{|c|c|c|c|c|c|}
        \hline
        Dataset & D30$\mu$K & W90 & W80 & W70 & W60\\
        \hline
        \makecell{1 (low-FG, \\$\beta=\alpha_i=0$)} & $0.02^\circ\pm0.19^\circ$ & $0.01^\circ\pm0.09^\circ$ & $0.02^\circ\pm0.13^\circ$ & $0.02^\circ\pm0.19^\circ$ & $0.02^\circ\pm0.20^\circ$ \\
        \hline
        \makecell{2 (high-FG, \\$\beta=\alpha_i=0$)} & $0.02^\circ\pm0.19^\circ$ & $0.02^\circ\pm0.09^\circ$ & $0.02^\circ\pm0.14^\circ$ & $0.01^\circ\pm0.20^\circ$ & $0.02^\circ\pm0.20^\circ$ \\
        \hline
        \makecell{3 (low-FG, \\$\beta=0.3^\circ$, $\alpha_i\neq0$)} & $0.27^\circ\pm0.19^\circ$ & $0.30^\circ\pm0.09^\circ$ & $0.31^\circ\pm0.13^\circ$ & $0.31^\circ\pm0.20^\circ$ & $0.31^\circ\pm0.19^\circ$ \\
        \hline
        \makecell{4 (high-FG, \\$\beta=0.3^\circ$, $\alpha_i\neq0$)} & $0.28^\circ\pm0.20^\circ$ & $0.31^\circ\pm0.09^\circ$ & $0.30^\circ\pm0.14^\circ$ & $0.31^\circ\pm0.20^\circ$ & $0.31^\circ\pm0.20^\circ$ \\
        \hline
    \end{tabular}
    \label{tab:ali-beta}
\end{table}

We show the measured parameters for Dataset 1 AliCPT combined with PR4 HFI simulations in Fig.~\ref{fig:ali-low-null}, and the corresponding uncertainties in Fig.~\ref{fig:sigma-ali-low-null}. We see that the wide-scan region masking the smallest Galactic sky, i.e., W90, results in the lowest uncertainty and smallest bias of measured $\beta$, which is $\hat{\beta}=0.00^\circ\pm0.09^\circ$. Despite the lower noise level, $\sigma(\beta)$ of the D30$\mu$K is larger than that of the wide-scan region due to lack of power to break the $\alpha_i+\beta$ degeneracy, which highlights the importance of a large sky fraction in birefringence estimation.

\revise{However, we note that the conclusion above is a consequence of the adopted methodology. Since the Minami-Komatsu method leverages the large-scale foreground observations to break the $\alpha_i+\beta$ degeneracy, it might be a suboptimal formalism for deep-scan scenarios given the small sky fraction. We will explore potentially superior alternatives (e.g., pixel-domain methods \cite{remazeillesFieldlevelConstraintsCosmic2025, Jost_2023}) in the future.}

The measured $\beta$ of all four datasets are listed in Tab.~\ref{tab:ali-beta}. The uncertainty of $\beta$ remains nearly unchanged across four datasets, indicating that the increase in complexity from the foreground and the rotated $E/B$ modes is subdominant in the covariance matrix. For all the masking cases of Dataset 4, our pipeline successfully recovers the input birefringence angle by $\hat{\beta}\approx0.3^\circ$, with the uncertainties varying from $0.09^\circ$ to $0.20^\circ$.

\begin{figure}[tbp]
    \centering
    \begin{subfigure}[b]{0.65\textwidth}
        \centering
        \includegraphics[width=\textwidth]{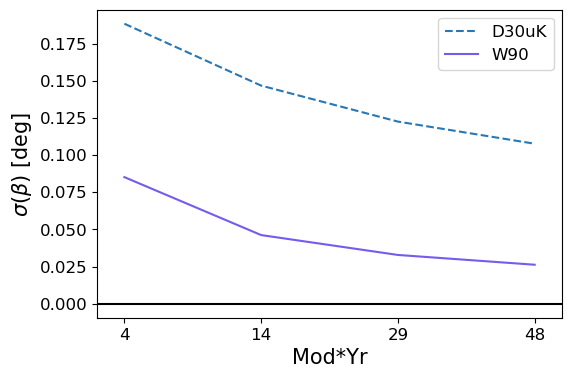}
    \end{subfigure}
    \caption{Sensitivities on AliCPT measurements of $\beta$ across different observing durations. The dashed and solid curves show the uncertainties of measured $\beta$ for D30$\mu$K and W90 masks, respectively.}
    \label{fig:forecast}
\end{figure}

The above results are based on the one-year AliCPT observations, whose ability to constrain the birefringence angle is similar to that of the \textit{Planck} nearly full-sky data (see Tab.~\ref{tab:hfi-beta}). We then forecast the sensitivities of $\beta$ corresponding to longer observing times of AliCPT. The number of AliCPT's modules and detectors increases year by year, and we use the accumulative product of the number of modules and the observing years, $N_{\rm mod}*N_{\rm yr}$, to describe the instrumental sensitivity of a period. After 1, 2, 3, 4 years' observations, we expect the $N_{\rm mod}*N_{\rm yr}$ of AliCPT to be 4, 14, 29, 48, respectively, with the decreasing noise level proportional to $[N_{\rm mod}*N_{\rm yr}]^{-1/2}$. We compute the uncertainties of $\beta$ from the Fisher matrix for these observing durations, as shown in Fig.~\ref{fig:forecast} and Tab.~\ref{tab:beta-duration}, using Dataset 2 simulations and two masks. We find the uncertainty could reach $0.026^\circ$ for W90 after four years' observation.

Finally we compare our results to the AliCPT paper \cite{liProbingPrimordialGravitational2019}. Ref. \cite{liProbingPrimordialGravitational2019} demonstrates that after 3 years' observations, AliCPT is capable of constraining $\beta$ at $\sim0.01^\circ$ level, without considering the foreground and instrumental effects. We run the estimator on the Dataset 2 simulations while fixing $\mathcal{A}=\alpha_i=0$, i.e., only fitting $\beta$. We find that under such case, after 3 years' observations, the AliCPT could reach $\sigma(\beta)=0.015^\circ$ and $0.012^\circ$ for D30$\mu$K and W90, respectively, which is consistent with the previous forecast. The comparison demonstrates that the foreground contamination and instrumental systematics significantly complicate the detection of the CB effect.

\begin{table}[tbp]
    \centering
    \caption{Uncertainties of measured $\beta$ for AliCPT+HFI simulations across different observing durations, computed from the Fisher matrix, assuming the high-complexity foreground.}
    \begin{tabular}{c|c|c|c}
        \hline
        \hline
        Year & $N_{\rm mod}*N_{\rm yr}$ & $\sigma(\beta)$ for D30$\mu$K & $\sigma(\beta)$ for W90 \\
        \hline
        1 & 4 & $0.19^\circ$ & $0.085^\circ$\\
        2 & 14 & $0.15^\circ$ & $0.046^\circ$\\
        3 & 29 & $0.12^\circ$ & $0.033^\circ$\\
        4 & 48 & $0.11^\circ$ & $0.026^\circ$\\
        \hline
        \hline
    \end{tabular}
    \label{tab:beta-duration}
\end{table}

\subsection{Impact of foregrounds}

\begin{figure}[tbp]
    \centering
    \begin{subfigure}[b]{0.9\textwidth}
        \centering
        \includegraphics[width=\textwidth]{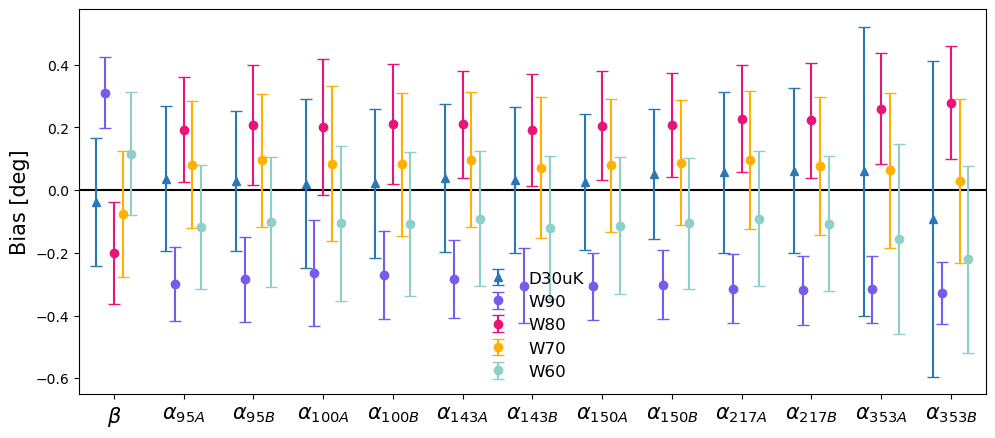}
    \end{subfigure}
    \caption{Bias of the measured parameters for AliCPT+HFI Dataset 4 (high-FG, $\beta=0.3^\circ$, $\alpha_i\neq0$) simulations while ignoring the foreground $EB$ spectrum in the likelihood.}
    \label{fig:ali-hig-rot-nofg}
\end{figure}

\begin{table}[tbp]
    \centering
    \caption{Bias of measured $\beta$ for AliCPT+HFI Dataset 2 simulations when ignoring the foreground $EB$ correlation.}
    \begin{tabular}{|c|c|c|c|c|c|}
        \hline
        Dataset & D30$\mu$K & W90 & W80 & W70 & W60\\
        \hline
        \makecell{1} & $0.00^\circ\pm0.21^\circ$ & $-0.44^\circ\pm0.13^\circ$ & $-0.02^\circ\pm0.16^\circ$ & $-0.05^\circ\pm0.20^\circ$ & $0.08^\circ\pm0.19^\circ$ \\
        \hline
        \makecell{2} & $0.03^\circ\pm0.20^\circ$ & $0.32^\circ\pm0.11^\circ$ & $-0.18^\circ\pm0.17^\circ$ & $-0.05^\circ\pm0.20^\circ$ & $0.15^\circ\pm0.19^\circ$ \\
        \hline
        \makecell{3} & $-0.06^\circ\pm0.21^\circ$ & $-0.45^\circ\pm0.13^\circ$ & $-0.04^\circ\pm0.16^\circ$ & $-0.07^\circ\pm0.20^\circ$ & $0.05^\circ\pm0.20^\circ$ \\
        \hline
        \makecell{4} & $-0.04^\circ\pm0.20^\circ$ & $0.31^\circ\pm0.11^\circ$ & $-0.20^\circ\pm0.16^\circ$ & $-0.08^\circ\pm0.20^\circ$ & $0.12^\circ\pm0.20^\circ$ \\
        \hline
    \end{tabular}
    \label{tab:ali-beta-ignfg}
\end{table}

To evaluate the impact of ignoring the foreground $EB$ spectrum, we test on the AliCPT+HFI simulations while fixing $\mathcal{A}=0$ in Eq.~\eqref{eq:llh}. We show the bias of measured $\beta$ in Tab.~\ref{tab:ali-beta-ignfg}, and the bias of all parameters for Dataset 4 in Fig.~\ref{fig:ali-hig-rot-nofg}. The high bias of the W90 mask (up to $3.5\sigma$) compared to other masks is because of its high amplitude of foregrounds under least masking, leading to a bias of $\gamma_\ell\approx C_\ell^{EB,\rm fg}/2(C_\ell^{EE,\rm fg}-C_\ell^{BB,\rm fg})$ on $\beta$ and $\alpha_i$ \cite{minamiNewExtractionCosmic2020,diego-palazuelosCosmicBirefringencePlanck2022}: $\hat{\beta}=\beta^{\rm true}-\gamma$, $\hat{\alpha}_i=\alpha_i^{\rm true}+\gamma$. This indicates that if we know nothing about the foreground $EB$ correlation, the D30$\mu$K mask with the lowest bias is a conservative choice. The results also show a clear discrepancy between the low- and high-complexity foregrounds for W90, due to the opposite sign of the $EB$ spectra between the two foreground templates.

Then we calculate the bias of parameters for Dataset 2 simulations (high-complexity foreground) while using the low-complexity foreground maps as the foreground templates. The bias of $\beta$ due to wrong foreground modeling is $0.04^\circ$ ($0.2\sigma$) and $0.35^\circ$ ($2.9\sigma$) for D30$\mu$K and W90 masks, respectively. The average of the fitted foreground amplitude $\mathcal{A}$ over simulations is 0.21 and 0.04 for D30$\mu$K and W90 masks, respectively, thus producing similar results to the $\mathcal{A}=0$ case.

\section{Conclusions}
\label{sec:conclusion}

We have presented forecasts for constraining isotropic cosmic birefringence using CMB simulations of the AliCPT experiment in combination with \textit{Planck} HFI data. Our analysis employs a semi-analytical maximum-likelihood framework that simultaneously estimates the birefringence angle $\beta$, instrumental miscalibration angles $\alpha_i$, and the amplitude of the foreground $EB$ spectrum $\mathcal{A}$. The method requires multi-frequency foreground templates as the prior of foreground $EB$ correlation.

The AliCPT telescope will implement two scanning scenarios: the wide scan with a higher sky coverage, and the deep scan with a lower noise level. We find that the wide-scan strategy of AliCPT with highest sky fraction of 38\% (e.g., W90 mask) yields the tightest constraint on $\beta$, with an uncertainty of $0.09^\circ$ for one-year AliCPT observations, significantly outperforming the deep-scan strategy ($\sigma(\beta)=0.20^\circ$) thanks to better degeneracy breaking between $\alpha_i$ and $\beta$. The uncertainties are robust against increases in complexity from the foreground model and CMB rotation. We also forecast that the uncertainty of $\beta$ will be further reduced to $0.026^\circ$ with four years of AliCPT data, indicating a $11\sigma$ detection of a birefringence angle $\beta=0.3^\circ$. These results demonstrate the AliCPT’s strong potential to detect or constrain cosmic birefringence, thereby probing parity-violating physics in the early universe.

Neglect or modeling imperfections of the foreground $EB$ correlation will introduce significant biases up to $3.5\sigma$ for the W90 mask, underscoring the necessity of accurate foreground modeling. However, using data of the deep-scan region with weak dust emissions could avoid such biases, offering a conservative choice given little knowledge of the foreground $EB$ correlation. In the case with real data, we should try all of these masking schemes to ensure their consistency and thereby verify the assumptions of our foreground model. \revise{Furthermore, as the Minami-Komatsu estimator relies on large-scale foreground observations to break the parameter degeneracy, we will explore more appropriate estimation frameworks for deep-scan strategies with limited sky coverage in the future.}

\appendix

\section{Calculation of the covariance matrix}
\label{app:der-cov}

In the following equations, we ignore terms containing $\sin(x) C_\ell^{EB}$ where $x$ is some small angle $\alpha_i$ or $\beta$, since these terms fluctuate rapidly around zero, making the covariance matrix unstable.

Following Eq.~\eqref{eq:cov_XYZW}, the covariance of the observed spectra is computed as:
\begin{equation}
\begin{aligned}
    \mathbf{C}^{\rm o*o}_{ijpqb}\approx\ & C_b^{E_i^{\rm o}E_p^{\rm o}}C_b^{B_j^{\rm o}B_q^{\rm o}}+C_b^{E_i^{\rm o}B_q^{\rm o}}C_b^{B_j^{\rm o}E_p^{\rm o}}\\
    &+\frac{s_{4\alpha_i}s_{4\alpha_q}}{[c_{4\alpha_i}+c_{4\alpha_j}][c_{4\alpha_p}+c_{4\alpha_q}]}\left(C_b^{E_i^{\rm o}E_p^{\rm o}}C_b^{E_j^{\rm o}E_q^{\rm o}}+C_b^{E_i^{\rm o}E_q^{\rm o}}C_b^{E_j^{\rm o}E_p^{\rm o}}\right)\\
    &+\frac{s_{4\alpha_i}s_{4\alpha_p}}{[c_{4\alpha_i}+c_{4\alpha_j}][c_{4\alpha_p}+c_{4\alpha_q}]}\left(C_b^{B_i^{\rm o}B_p^{\rm o}}C_b^{B_j^{\rm o}B_q^{\rm o}}+C_b^{B_i^{\rm o}B_q^{\rm o}}C_b^{B_j^{\rm o}B_p^{\rm o}}\right)\,.
\end{aligned}
\end{equation}
Note that although the auto-frequency observed spectra with $i=j$ or $p=q$ are excluded from the likelihood (Eq.~\eqref{eq:llh}), they still appear in the computation of the covariance matrix with $i=p$ or $j=q$ since indeed, the noise bias should be taken into account in the covariance. So all the auto-frequency spectra must be calculated before the iterative process.

The covariance of the foreground templates is written as:
\begin{equation}
\begin{aligned}
    \mathbf{C}^{\rm fg*fg}_{ijpqb}\approx\ & \frac{4c_{2\alpha_i}c_{2\alpha_j}c_{2\alpha_p}c_{2\alpha_q}}{[c_{4\alpha_i}+c_{4\alpha_j}][c_{4\alpha_p}+c_{4\alpha_q}]}\mathcal{A}^2\left(C_b^{E_i^{\rm fg}E_p^{\rm fg}}C_b^{B_j^{\rm fg}B_q^{\rm fg}}+C_b^{E_i^{\rm fg}B_q^{\rm fg}}C_b^{B_j^{\rm fg}E_p^{\rm fg}}\right)\\
    &+\frac{4c_{2\alpha_i}c_{2\alpha_j}s_{2\alpha_p}s_{2\alpha_q}}{[c_{4\alpha_i}+c_{4\alpha_j}][c_{4\alpha_p}+c_{4\alpha_q}]}\mathcal{A}^2 C_b^{E_i^{\rm fg}E_q^{\rm fg}}C_b^{B_j^{\rm fg}B_p^{\rm fg}}\\
    &+\frac{4s_{2\alpha_i}s_{2\alpha_j}c_{2\alpha_p}c_{2\alpha_q}}{[c_{4\alpha_i}+c_{4\alpha_j}][c_{4\alpha_p}+c_{4\alpha_q}]}\mathcal{A}^2 C_b^{B_i^{\rm fg}B_q^{\rm fg}}C_b^{E_j^{\rm fg}E_p^{\rm fg}}\\
    &+\frac{4s_{2\alpha_i}s_{2\alpha_j}s_{2\alpha_p}s_{2\alpha_q}}{[c_{4\alpha_i}+c_{4\alpha_j}][c_{4\alpha_p}+c_{4\alpha_q}]}\mathcal{A}^2 C_b^{B_i^{\rm fg}B_p^{\rm fg}}C_b^{E_j^{\rm fg}E_q^{\rm fg}}\,.
\end{aligned}
\end{equation}

The cross correlation between the observed signal and the foreground templates is given by:
\begin{equation}
\begin{aligned}
    \mathbf{C}^{\rm fg*o}_{ijpqb}\approx\ & \frac{2c_{2\alpha_i} c_{2\alpha_j}}{c_{4\alpha_i}+c_{4\alpha_j}}\mathcal{A}\left(C_b^{E_i^{\rm fg}E_p^{\rm o}}C_b^{B_j^{\rm fg}B_q^{\rm o}}+C_b^{E_i^{\rm fg}B_q^{\rm o}}C_b^{B_j^{\rm fg}E_p^{\rm o}}\right)\\
    &+\frac{2c_{2\alpha_p} c_{2\alpha_q}}{c_{4\alpha_p}+c_{4\alpha_q}}\mathcal{A}\left(C_b^{E_i^{\rm o}E_p^{\rm fg}}C_b^{B_j^{\rm o}B_q^{\rm fg}}+C_b^{E_i^{\rm o}B_q^{\rm fg}}C_b^{B_j^{\rm o}E_p^{\rm fg}}\right)\\
    &+\frac{2s_{2\alpha_i} s_{2\alpha_j}}{c_{4\alpha_i}+c_{4\alpha_j}}\mathcal{A} C_b^{B_i^{\rm fg}B_q^{\rm o}}C_b^{E_j^{\rm fg}E_p^{\rm o}}\\
    &+\frac{2s_{2\alpha_p} s_{2\alpha_q}}{c_{4\alpha_p}+c_{4\alpha_q}}\mathcal{A} C_b^{E_i^{\rm o}E_q^{\rm fg}}C_b^{B_j^{\rm o}B_p^{\rm fg}}\,.
\end{aligned}
\end{equation}

The full covariance matrix is finally computed as Eq.~\eqref{eq:cov_full}.

\section{Linear equations of the maximum likelihood solution}
\label{app:lin-eqs}

The diagonal elements of the symmetric matrix $\mathbf{A}_{mn}$ are:
\begin{equation}
\begin{aligned}
    \mathbf{A}_{\rm fg,fg}=&\sum_{\substack{b\\i,j\neq i\\p,q\neq p}} C_b^{E_i^{\rm fg}B_j^{\rm fg}} \mathbf{C}^{-1}_{ijpqb} C_b^{E_p^{\rm fg}B_q^{\rm fg}}\,,\\
    \mathbf{A}_{\beta,\beta}=&\ 4\sum_{\substack{b\\i,j\neq i\\p,q\neq p}} \left(C_b^{E_i^{\rm CMB}E_j^{\rm CMB}}-C_b^{B_i^{\rm CMB}B_j^{\rm CMB}}\right) \mathbf{C}^{-1}_{ijpqb} \left(C_b^{E_p^{\rm CMB}E_q^{\rm CMB}}-C_b^{B_p^{\rm CMB}B_q^{\rm CMB}}\right)\,,\\
    \mathbf{A}_{\alpha_m,\alpha_n}=&\ 4\sum_{\substack{b\\i\neq n\\j\neq m}} \left(C_b^{E_i^{\rm o}E_n^{\rm o}} \mathbf{C}^{-1}_{injmb} C_b^{E_j^{\rm o}E_m^{\rm o}}+C_b^{B_n^{\rm o}B_i^{\rm o}} \mathbf{C}^{-1}_{nimjb} C_b^{B_m^{\rm o}B_j^{\rm o}}\right)\\
    &-4\sum_{\substack{b\\i\neq n\\j\neq m}} \left(C_b^{B_n^{\rm o}B_i^{\rm o}} \mathbf{C}^{-1}_{nijmb} C_b^{E_j^{\rm o}E_m^{\rm o}}+C_b^{B_m^{\rm o}B_j^{\rm o}} \mathbf{C}^{-1}_{mjinb} C_b^{E_i^{\rm o}E_n^{\rm o}}\right)\,,
\end{aligned}
\end{equation}
and the off-diagonal elements are:
\begin{equation}
\begin{aligned}
    \mathbf{A}_{\rm{fg},\beta}=&\ 2\sum_{\substack{b\\i,j\neq i\\p,q\neq p}} C_b^{E_i^{\rm fg} B_j^{\rm fg}}\mathbf{C}^{-1}_{ijpqb}\left(C_b^{E_p^{\rm CMB} E_q^{\rm CMB}}-C_b^{B_p^{\rm CMB} B_q^{\rm CMB}}\right)\,,\\
    \mathbf{A}_{\rm{fg},\alpha_n}=&\ 2\sum_{\substack{b\\i\neq n\\p,q\neq p}} \left(C_b^{E_i^{\rm o} E_n^{\rm o}}\mathbf{C}^{-1}_{inpqb}C_b^{E_p^{\rm fg} B_q^{\rm fg}}-C_b^{B_n^{\rm o} B_i^{\rm o}}\mathbf{C}^{-1}_{nipqb}C_b^{E_p^{\rm fg} B_q^{\rm fg}}\right)\,,\\
    \mathbf{A}_{\beta,\alpha_n}=&\ 4\sum_{\substack{b\\i\neq n\\p,q\neq p}}\left(C_b^{E_i^{\rm o} E_n^{\rm o}}\mathbf{C}^{-1}_{inpqb}-C_b^{B_n^{\rm o} B_i^{\rm o}}\mathbf{C}^{-1}_{nipqb}\right)\left(C_b^{E_p^{\rm CMB} E_q^{\rm CMB}}-C_b^{B_p^{\rm CMB} B_q^{\rm CMB}}\right)\,.
\end{aligned}
\end{equation}
The elements of the vector $\mathbf{b}_{m}$ are:
\begin{equation}
\begin{aligned}
    \mathbf{b}_{\rm fg}=&\sum_{\substack{b\\i,j\neq i\\p,q\neq p}} C_b^{E_i^{\rm o}B_j^{\rm o}} \mathbf{C}^{-1}_{ijpqb} C_b^{E_p^{\rm fg}B_q^{\rm fg}}\,,\\
    \mathbf{b}_{\beta}=&\ 2\sum_{\substack{b\\i,j\neq i\\p,q\neq p}} C_b^{E_i^{\rm o}B_j^{\rm o}} \mathbf{C}^{-1}_{ijpqb} \left(C_b^{E_p^{\rm CMB}E_q^{\rm CMB}}-C_b^{B_p^{\rm CMB}B_q^{\rm CMB}}\right)\,,\\
    \mathbf{b}_{\alpha_m}=&\ 2\sum_{\substack{b\\i\neq m\\p,q\neq p}}\left(C_b^{E_i^{\rm o} E_m^{\rm o}}\mathbf{C}^{-1}_{impqb}C_b^{E_p^{\rm o} B_q^{\rm o}}-C_b^{B_m^{\rm o} B_i^{\rm o}}\mathbf{C}^{-1}_{mipqb}C_b^{E_p^{\rm o} B_q^{\rm o}}\right)\,.
\end{aligned}
\end{equation}






\section{Selection of the multipole range}
\label{app:vary-lrange}

\begin{figure}[tbp]
    \centering
    \begin{subfigure}[b]{0.9\textwidth}
        \centering
        \includegraphics[width=\textwidth]{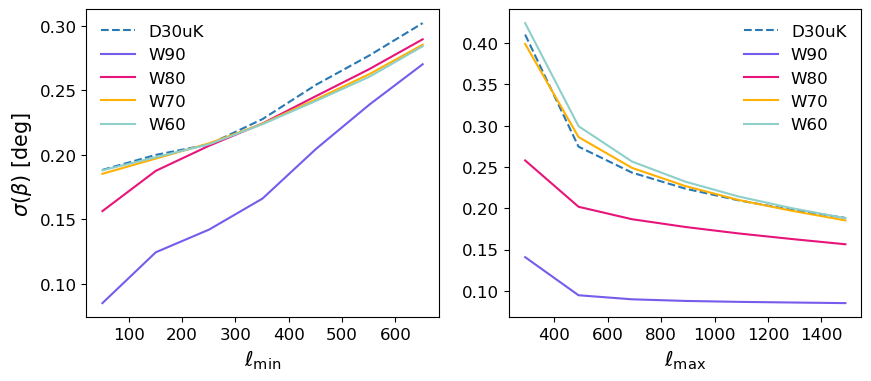}
    \end{subfigure}
    \caption{Uncertainties of measured $\beta$ for Dataset 2 simulations when changing the multipole range. \textit{Left}: uncertainties for the multipole range $\ell\in[\ell_{\min}, 1490]$. \textit{Right}: uncertainties for $\ell\in[51, \ell_{\max}]$.}
    \label{fig:vary-lrange}
\end{figure}

We adjust the multipole range to determine the optimal one for AliCPT+HFI simulations. The uncertainty of measured $\beta$ is expected to decrease with a wider multipole range, but the signal-to-noise ratio (S/N) of $\beta$ might saturate under a narrower range. We show the uncertainties of $\beta$ for Dataset 2 varying either $\ell_{\min}$ or $\ell_{\max}$ in Fig.~\ref{fig:vary-lrange}. Although the uncertainty does not saturate at $\ell_{\min}=51$, the measured $\beta$ with lower $\ell_{\min}$ would be unstable due to non-Gaussian effects from e.g., foregrounds \cite{minamiNewExtractionCosmic2020}. When varying $\ell_{\max}$, the uncertainty nearly converge at $\ell_{\max}\approx600$ for W90, while for D30$\mu$K and W70 the uncertainty does not converge. Hence we still use $\ell\in[51, 1490]$ as the baseline multipole range for AliCPT+HFI. The turning point at $\ell_{\max}\approx500$ on the right panel corresponds to the trough of the birefringence power spectrum $C_\ell^{EE,\rm CMB}-C_\ell^{BB,\rm CMB}$, beyond which the growth of the S/N slows down.

\section{\revise{Validation of the use of data splits}}
\label{app:valid_data_split}

In Sec.~\ref{sec:form_eb_sp}, we mentioned that we divided the two AliCPT bands into A/B data splits and excluded auto-spectra to avoid the correlated systematics and noise bias. Here we will show that this approach would not lose any precision, compared with using the full-mission maps or including the auto-spectra. For Dataset 1 simulations, we implement four AliCPT data configurations as follows:
\begin{enumerate}
    \item Fit $\alpha_{\rm 95A}$, $\alpha_{\rm 95B}$, $\alpha_{\rm 150A}$ and $\alpha_{\rm 150B}$ from cross-power spectra between data splits, excluding auto-power spectra (e.g., $\rm 95A\times 95A$). This is the baseline configuration in the main body of this work.
    \item Fit $\alpha_{\rm 95A}$, $\alpha_{\rm 95B}$, $\alpha_{\rm 150A}$ and $\alpha_{\rm 150B}$ from cross-power spectra between data splits but including auto-spectra of four 95A, 95B, 150A and 150B splits.
    \item Fit $\alpha_{\rm 95}$ and $\alpha_{\rm 150}$ from $\rm 95\times 150$ cross-power spectra between 95 and 150 GHz full-mission maps.
    \item Fit $\alpha_{\rm 95}$ and $\alpha_{\rm 150}$ from auto- and cross-power spectra between full-mission maps, that is, including $\rm 95\times 95$, $\rm 95\times 150$, and $\rm 150\times 150$ spectra.
\end{enumerate}
The PR4 HFI configuration remains the same across all four cases: we fit the miscalibration angle for each A/B data split of HFI bands independently, excluding the auto-data-split spectra.

\begin{figure}[tbp]
    \centering
    \begin{subfigure}[b]{0.9\textwidth}
        \centering
        \includegraphics[width=\textwidth]{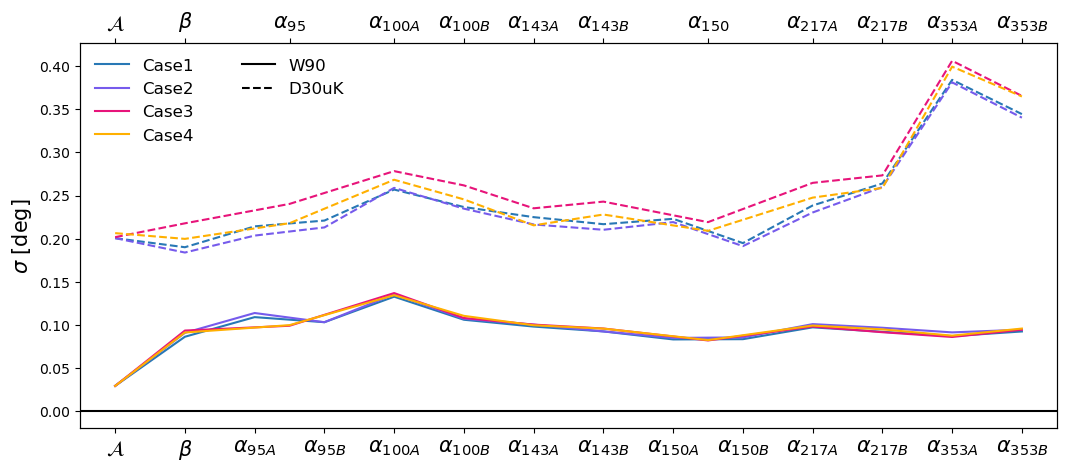}
    \end{subfigure}
    \caption{Simulations' dispersion of the measured parameters across four AliCPT cases for Dataset 1 simulations, using D30$\mu$K (dashed) and W90 (solid) masks. The lower $x$-axis is for Case 1 and 2, where we fit angles for each data split independently, while the upper $x$-axis is for Case 3 and 4, where we fit one miscalibration angle for AliCPT 95 or 150 GHz band.}
    \label{fig:valid_ds}
\end{figure}

\begin{table}[tbp]
    \centering
    \caption{Measured $\beta$ for Dataset 1 simulations using four AliCPT data configurations. The values in parentheses are the Fisher uncertainties.}
    \begin{tabular}{c|c|c}
        \hline
        \hline
        Case & D30$\mu$K & W90 \\
        \hline
        1 (data split, cross-only) & $0.02^\circ\pm0.19^\circ\ (0.19^\circ)$ & $0.01^\circ\pm0.09^\circ\ (0.10^\circ)$ \\
        2 (data split, with auto) & $0.01^\circ\pm0.18^\circ\ (0.18^\circ)$ & $0.01^\circ\pm0.09^\circ\ (0.10^\circ)$ \\
        3 (full-mission, cross-only) & $0.00^\circ\pm0.22^\circ\ (0.22^\circ)$ & $0.00^\circ\pm0.09^\circ\ (0.11^\circ)$ \\
        4 (full-mission, with auto) & $0.00^\circ\pm0.20^\circ\ (0.21^\circ)$ & $0.00^\circ\pm0.09^\circ\ (0.10^\circ)$ \\
        \hline
        \hline
    \end{tabular}
    \label{tab:valid_ds}
\end{table}

The results of the four cases applied to Dataset 1 simulations, using D30$\mu$K and W90 masks are shown in Fig.~\ref{fig:valid_ds} and Tab.~\ref{tab:valid_ds}. We find that all the cases render similar uncertainties of parameters for W90, while for D30$\mu$K the uncertainty of $\beta$ in Case 3 is slightly larger than others, since Case 3 utilizes least information of power spectra among four cases. Even though the full-mission case (Case 4) fits two fewer parameters, it does not outperform the data-split case (Case 2). Because of the strong positive correlation between two miscalibration angles within a single band, the increase in the parameter space does not substantially impact the uncertainty of $\beta$. The inclusion of auto-spectra yields little improvement in the sensitivity to $\beta$, which is consistent with the findings for \textit{Planck} HFI-only simulations, as reported in \cite{diego-palazuelosRobustnessCosmicBirefringence2023}.

The ACT team divide the DR6 time streams into four independent time chunks (data splits), and calculate the final unbiased power spectra as the average across all cross-data-split power spectra for each frequency band \cite{ACT:2025fju}. However, this approach becomes computationally demanding when applied to forecasts involving a large number of simulations. We therefore suggest it as a potential improved avenue for future work.

\acknowledgments
JD thanks Patricia Diego-Palazuelos for useful discussions and comments.
This work is supported by NSFC No. 12325301 and 12273035, the National Key R\&D Program of China Grant No. 2021YFC2203102 and 2022YFC2204602, Strategic Priority Research Program of the Chinese Academy of Science Grant No. XDB0550300.

\bibliography{Bib}
\bibliographystyle{JHEP}

\end{document}